\documentclass[
reprint,
superscriptaddress,
amsmath,
amssymb,
longbibliography,twocolumn
]{revtex4-2}

\usepackage{amsmath}
\usepackage{mathtools}
\usepackage{amssymb}
\usepackage{graphicx}
\usepackage{bm}
\usepackage[colorlinks, linkcolor=blue, citecolor=blue, urlcolor=blue,breaklinks=true]{hyperref}
\usepackage{color,dsfont,multirow,booktabs}
\usepackage{braket}
\usepackage{tabularx}
\usepackage{lipsum,booktabs}

\usepackage{ulem}

\usepackage[table]{xcolor}
\definecolor{boxblue}{rgb}{0.90, 0.94, 1.0}

\definecolor{boxheadblue}{rgb}{0.70, 0.85, 1.0}

\begin{document}
\title{Nonreciprocity in Quantum Technology}

\author{Shabir Barzanjeh}
\email{shabir.barzanjeh@ucalgary.ca}
\affiliation{Institute for Quantum Science and Technology, and Department of Physics and Astronomy University of Calgary,
2500 University Drive NW, Calgary, Alberta T2N 1N4, Canada}
\author{Andr\'e Xuereb}
\affiliation{Department of Physics, University of Malta, Msida MSD\,2080, Malta.}
\author{Andrea Al\`u}
\affiliation{Photonics Initiative, Advanced Science Research Center, City University of New York, New York, NY 10031, USA.}
\affiliation{Physics Program, Graduate Center, City University of New York, New York, NY 10016, USA.}
\author{Sander A.\ Mann}
\affiliation{Institute of Physics, University of Amsterdam, Amsterdam, The Netherlands.}
\author{Nikita Nefedkin}
\affiliation{Physics Program, Graduate Center, City University of New York, New York, NY 10016, USA.}
\author{Vittorio Peano}
\affiliation{Max Planck Institute for the Science of Light, Staudtstraße 2, 91058 Erlangen, Germany.}
\author{Peter Rabl}
\affiliation{Technical University of Munich, TUM School of Natural Sciences, Physics Department, 85748 Garching, Germany.}
\affiliation{Walther-Meißner-Institut, Bayerische Akademie der Wissenschaften, 85748 Garching, Germany.}
\affiliation{Munich Center for Quantum Science and Technology (MCQST), 80799 Munich, Germany.}

\begin{abstract}

Nonreciprocity—the ability to transmit signals in one direction while blocking them in the reverse—has become a powerful resource in quantum technologies, enabling directional amplification, routing of quantum information, and topologically protected quantum states. Recent experimental advances have demonstrated nonreciprocal behavior in low-loss, fully integrated devices operating with weak or no magnetic bias, enabled by synthetic gauge fields, optomechanical interactions, and chiral light–matter coupling. These achievements overcome the limitations of more traditional approaches, making nonreciprocity compatible with superconducting circuits and scalable quantum photonic architectures as well as an integral part of the next generation of modular quantum computers, distributed quantum networks, and precision metrology. Here we highlight the key concepts for engineering nonreciprocity in quantum systems and describe how this functionality can be employed for high-fidelity qubit readout, robust quantum state transfer, and boosting the sensitivity of quantum sensors. 
\end{abstract}
\maketitle

Reciprocity is the principle that wave transmission remains invariant when the roles of source and receiver are exchanged. As such, it is a fundamental property of linear, time-reversal symmetric systems \cite{Potton2004, Deak2012}. Techniques for breaking this symmetry \cite{Jalas2013, Caloz2018} are of immediate relevance for signal transmission and sensing applications; nonreciprocal components enable, for example, the isolation of sources from detectors to minimise unwanted feedback or directional signal routing. There has recently been growing interest in extending nonreciprocity into the quantum regime, enabling unidirectional quantum transport~\cite{Yao2013,Lemonde2019,Ownes2022,Lodahl2017,DeBernardis2023},
quantum-limited amplification \cite{Kamal2011}, enhanced sensing \cite{Lau2018, Chen_2019, McDonald2020}, and topological protection \cite{Wang2009,Hafezi2011,Ringel2014,Peano2015,Peano_2016,Ozawa2019}. These functionalities are particularly critical for quantum communication and quantum information processing, where precise control of information flow, noise suppression, and preservation of coherence are essential requirements. Advances in scalable quantum technologies are therefore deeply intertwined with the development of nonreciprocal quantum components of equally high fidelity.  

Traditional approaches to nonreciprocity primarily rely on effects such as Faraday rotation, where an applied magnetic field breaks time-reversal symmetry \cite{Jalas2013}. While similar concepts can be applied at the quantum level by exploiting magnonic degrees of freedom \cite{PhysRevLett.123.127202,PhysRevA.101.043842,PhysRevApplied.13.044039, Kim2024, Ownes2022}, material losses, bulky designs or detrimental stray magnetic fields in superconducting platforms pose challenges for their large-scale integration. These limitations have driven increasing interest in alternative  mechanisms that are capable of achieving equivalent nonreciprocal responses without magnetic fields. These approaches typically break time-reversal symmetry in nonlinear systems, where a directional signal flow is imposed by strong external driving fields~\cite{Sounas2017} without the need for a magnetic bias.
Indeed, recent experiments have demonstrated that nonreciprocity can be realized via synthetic gauge fields \cite{Estep2014, Tzuang2014, Fang2012,Roushan2017,Rosen2024}, nonlinearities \cite{regensburger2012,miri2019, hua2016, huang2021}, optomechanical interactions \cite{Barzanjeh2017,Bernier2017, Malz2018,Mirhosseini2020}, and parametric modulation \cite{PhysRevLett.113.247003, Kamal2011}.  Such approaches are especially attractive for on-chip quantum processors and integrated photonic circuits, as they can be compact and avoid stray magnetic fields.

In this Review, we provide a comprehensive overview of recent progress in nonreciprocal quantum systems and their quantum technological applications. Our discussion covers a range of nonreciprocal platforms that have emerged, including optomechanical systems \cite{Aspelmeyer2014, Barzanjeh2022}, superconducting circuits \cite{Blais2004, GU20171, RevModPhys.93.025005}, and chiral atom-photon interfaces \cite{Lodahl2017} as prominent examples. In these systems, engineered nonreciprocal interactions enable the routing and faithful transmission of quantum states, directional amplification, enhanced quantum sensing, and the simulation of quantum many-body systems in the presence of effective magnetic fields. In what follows, we highlight both the theoretical concepts and current experimental approaches that underlie these transformative applications of quantum nonreciprocity.
\begin{figure}[ht!]
\centering{\includegraphics[width=1\columnwidth]{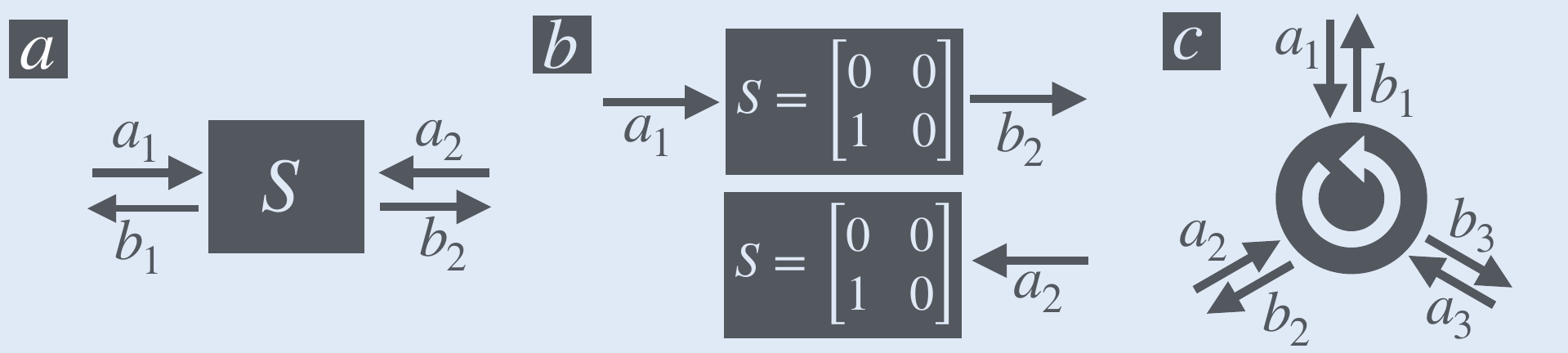}
     \caption{\textbf{Nonreciprocal devices.}
    \textbf{a}, A two-port device can be represented mathematically by its scattering matrix, $S$. \textbf{b}, An optical isolator that transmit signals in one direction but not in the other. \textbf{c}, Sketch of a three-port circulator. }}
     \label{boxschematic}
\end{figure}

\noindent \fcolorbox{black}{boxblue}{
  \begin{minipage}{1\columnwidth}
\textbf{Box 1| Basic principles of nonreciprocity}
\vspace{3mm}

For a linear two-port device as shown in Fig.\ \ref{boxschematic}\textbf{a}, the outgoing signals $b_i$ in ports $i=1,2$ are related to the corresponding input signals $a_i$ by a linear scattering relation of the form
\begin{align}
\begin{pmatrix} b_1 \\ b_2 \end{pmatrix} = 
\begin{bmatrix} S_{11} & S_{12} \\ S_{21} & S_{22} \end{bmatrix}  
\begin{pmatrix} a_1 \\ a_2 \end{pmatrix}.
\end{align}
The elements \( S_{ij} \) of the scattering matrix (or \textit{S-matrix}) describe how signals propagate between ports $i,j=1,2$. 

In a reciprocal system, the S-matrix is fully symmetric, meaning that $S_{12} = S_{21}$. This implies that a signal travelling from port 1 to port 2 behaves identically to one travelling from 2 to 1. In a nonreciprocal system, this symmetry is broken and $S_{12} \neq S_{21}$. An extreme example is an optical isolator, for which
\begin{align}
   S= \begin{bmatrix} 0 & 0 \\ 1 & 0 \end{bmatrix},
\end{align}
which models perfect transmission of signals from port 1 to port 2, with transfer in the opposite direction blocked, as shown in Fig.\ \ref{boxschematic}\textbf{b}. While this S-matrix describes a dissipative device where energy flowing from port 2 to 1 is fully absorbed, nonreciprocity could also exist in energy conserving systems. A prototypical example is a three-port circulator with S-matrix 
\begin{align}
   S= \begin{bmatrix} 0 & 0 & 1 \\ 1 & 0 & 0 \\
   0 &1  & 0 \end{bmatrix}.
\end{align}
Here signals are redirected cyclically ($a_1 \rightarrow b_{2}$, $a_2 \rightarrow b_{3}$, and $a_3 \rightarrow b_{1}$) in an energy-conserving manner. This device can be converted into a two-port isolator, for example, by terminating one of the ports by an absorbing medium.
\end{minipage}
}

\section*{Breaking reciprocity}
Nonreciprocity (see Box~1) can be best understood in terms of the scattering matrix $S$, which relates the input and output signals of a device and satisfies $S=S^T$ for a reciprocal system. A basic example, where this condition is broken, is an isolator, which allows signals to pass from one port to another but blocks any reverse transmission. Another common nonreciprocal device is the circulator,  which routes signals cyclically among three ports. This process can occur without energy loss, which makes it particularly relevant for coherent quantum applications.  However, according to the Lorentz reciprocity theorem, the response of nonmagnetic, time-invariant, linear materials---as primarily used for signal transmission---is fully symmetric, even in the presence of gain or loss~\cite{Jalas2013, Caloz2018}. Therefore, strategies to realize a nonreciprocal response must overcome this fundamental constraint. In quantum systems, this may be achieved in several ways, including by (I) using magnetic materials, (II) introducing parametric driving, (III) preparing the device in a quantum state with broken time-reversal symmetry, (IV) introducing strong few-photon nonlinearities.

\begin{figure}[t!]
\centering{\includegraphics[width=1\columnwidth]{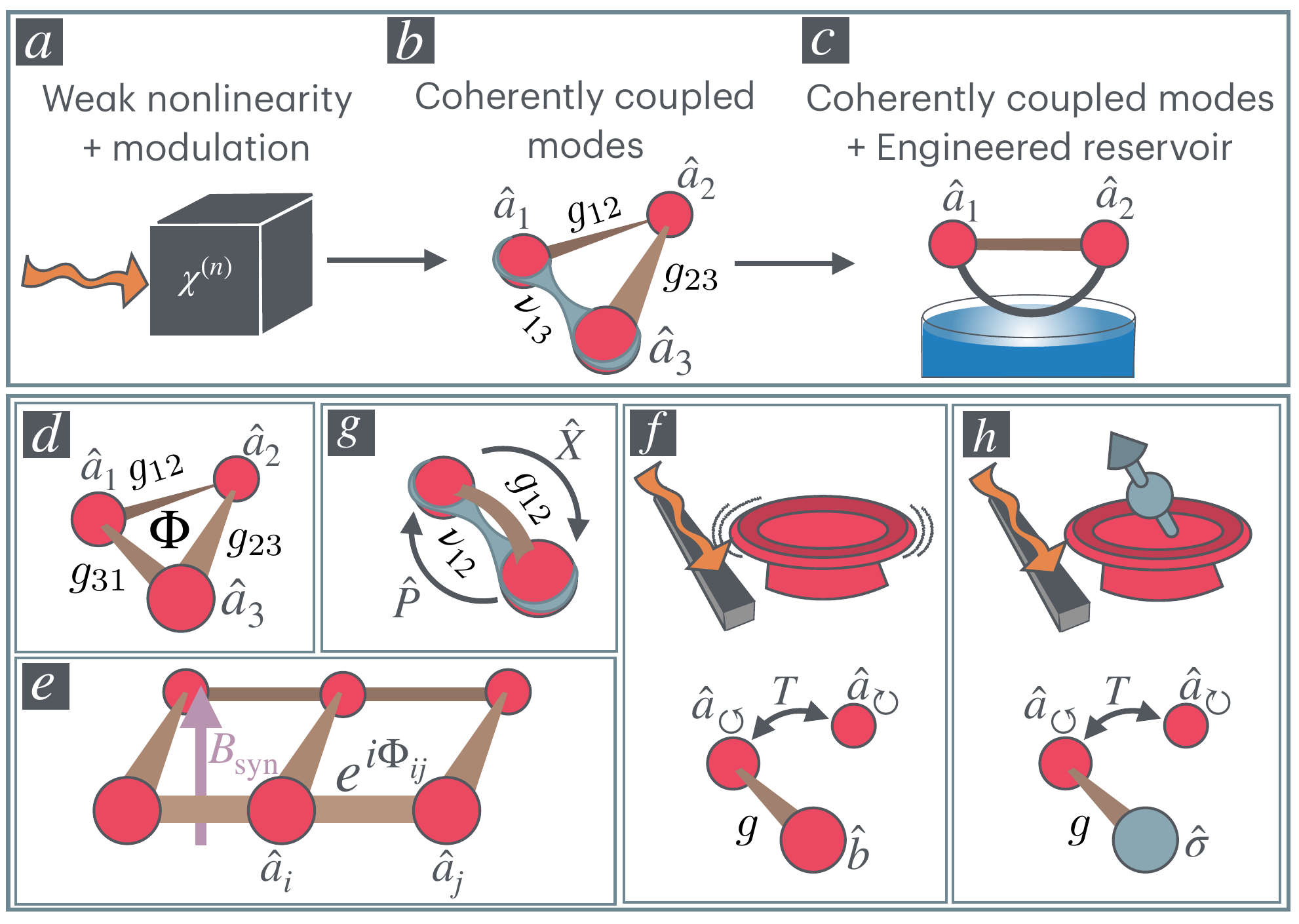}}
\caption{\textbf{Mechanisms to break reciprocity.}
    \textbf{a}, A common route to nonreciprocity, which is compatible with on-chip low-noise operations, consists in modulating a weakly nonlinear system.
    \textbf{b}, The graph representation of the linearized effective system. Each vertex represents a mode, and each edge a driving-induced coherent quadratic coupling. Brown and blue edges correspond to beam-splitting and two-mode squeezing coupling, respectively; the labels display the coupling constants. \textbf{c}, Equivalent reservoir engineering model when one mode decays rapidly. A reservoir mediating a dissipative coupling (black edge) replaces the fast-decaying mode $\hat{a}_3$.
    \textbf{d}–\textbf{h}, Examples of quantum networks leading to a nonreciprocal response:
    \textbf{d}, Graph representation of a circulator. Each mode is coupled in reflection to a waveguide (not shown). The time-reversal symmetry is broken by the artificial magnetic flux $\Phi=\arg (g_{12}g_{23}g_{31})$.
    \textbf{e}, This scheme can be generalized to a whole lattice of modes to simulate photonic systems with a synthetic magnetic field $B_{\rm syn}$. 
    \textbf{f}, Optomechanical isolator. A microtoroid supports pairs of whispering gallery modes ($\hat{a}_\circlearrowleft$ and $\hat{a}_\circlearrowright$). They are weakly coupled to a breathing mechanical mode $\hat{b}$. The directional drive (orange arrow) selectively pumps $\hat{a}_\circlearrowleft$. Time-reversal symmetry is broken because $\hat{b}$ is coupled to $\hat{a}_\circlearrowleft$ but not to $\hat{a}_\circlearrowright$. 
    \textbf{g}, Graph representation of two-mode nonreciprocal device with quadrature nonreciprocity. The parameters can be chosen such that the excitations on different conjugated quadratures ($\hat{X}$ and  $\hat{P}$) move in opposite directions (as indicated by the black arrows).
    \textbf{h}, Hybrid chiral isolator. An atom is coupled via polarization-sensitive transitions to a microtoroid. After the directional drive prepares the atom in a Zeeman state, only the mode $\hat{a}_\circlearrowleft$  is effectively coupled to the atom via the relevant transition operator $\hat{\sigma}$.
\label{fig_mechanims}}
\end{figure}
\textit{(I) Magnetic quantum systems.}---Similar to classical devices, reciprocity can be broken in quantum systems that involve magnetic degrees of freedom. A common choice is to use collective spin waves (magnons) in ferromagnetic materials such as yttrium iron garnet (YIG) \cite{PhysRevLett.123.127202, PhysRevA.101.043842,PhysRevApplied.13.044039, Kim2024} which interact strongly with microwave photons when embedded in a cavity. In a cylindrical geometry,  the left- and right-circulating microwave modes couple to a common magnon mode (e.g., the Kittel mode) with different strengths, resulting in a nonreciprocal behavior of the hybridized magnon--photon modes. A drawback of this scheme is that these hybridized excitations inherit the intrinsic losses of the magnons, which are fundamentally limited by the material properties rather than by design.

\textit{(II) Parametrically driven quantum systems.}---A general approach to introduce nonreciprocity consists in driving a weakly-nonlinear device, as depicted in Fig.~\ref{fig_mechanims}\textbf{a}. One or more driving fields could be injected directly from the ports of an otherwise reciprocal device. The nonlinearity is required to introduce parametric driving, which in turn generally breaks the time-invariance and time-reversal symmetries. The required nonlinearity could arise, e.g., from the Pockels or Kerr effect in (electro-)optical systems, radiation pressure in opto- or electromechanics, or the Josephson junction nonlinearity in superconducting microwave circuits. The basic principle  can be illustrated in terms of the $\chi ^{(2)}$-type nonlinear interaction $\hat{H}_{\chi}= \chi ( \hat a_1 \hat a_2^\dag \hat b + \hat  b^\dag  \hat  a_1^\dag \hat  a_2)$, which couples the cavity modes $\hat  a_{1}$ and $\hat  a_{2}$ via an auxiliary mode $\hat b$; other interactions, including $\chi ^{(3)}$-type, could similarly be used. Under strong driving, the auxiliary mode can be replaced by its mean amplitude, $\hat b\rightarrow \langle \hat b\rangle \in C$, giving rise to an effective \textit{beam-splitting} interaction $ \hat{H}_{\chi} \simeq g_{12} \hat{a}_1^\dagger \hat{a}_2+\text{h.c.}$ for the remaining modes of interest. The key observation is that both the magnitude and the phase of the resulting coupling constant, $g_{12}=\chi\langle \hat b\rangle$, are controlled by the external driving field.  Similar driving schemes that modulate the coupling between two modes close to the sum of their resonance frequencies can be employed to realize \textit{two-mode squeezing} interactions $ \nu_{12} \hat{a}_1^\dagger \hat{a}^\dagger_2+\text{h.c.}$ with tunable $\nu_{12} \in C$. Such squeezing interactions give rise to the creation of entangled photon pairs and to parametric amplification. As depicted in Fig.~\ref{fig_mechanims}\textbf{b}, when generalized to multiple modes, this approach enables the realization of arbitrarily coupled linear networks of bosonic modes, which generically result in nonreciprocal scattering relations with respect to input and output ports. To gain insight on the network response it is important to notice that, in the ideal limit of overcoupled ports, the scattering matrix is fully determined by the dimensionless parameters $g_{ij}\big/\sqrt{\kappa_i\kappa_j}$ and $\nu_{ij}\big/\sqrt{\kappa_i\kappa_j}$, where $g_{ii}$ is the detuning of mode $i$ from the carrier frequency and $\kappa_i$ denotes the decay rate into an intrinsic loss channel or the outcoupling into a device port. This lack of explicit dependence on the decay rates reduces the number of free parameters; leveraging this simplification, Landgraf, \textit{et al.}, \cite{Landgraf_2025, github_autoscattering} developed an algorithm that automatically discovers all network configurations implementing any target ideal nonreciprocal device.

Another way to gain physical insight is to interpret the fast-decaying modes as nonlocal reservoirs that mediate purely dissipative nonreciprocal interactions \cite{Metelmann2015, Metelmann2017}, as shown in Fig.~\ref{fig_mechanims}\textbf{c}. Alternatively, the dynamical matrix that governs the deterministic (drift) part of the dissipative dynamics can be interpreted as a non-Hermitian Hamiltonian\cite{Wang_2019}. Next, we discuss some general strategies for implementing nonreciprocity in driven quantum systems.

\textit{(IIa) Synthetic gauge fields.}---In the first scenario (Fig.~\ref{fig_mechanims}\textbf{d}),  an excitation moving on a closed loop from mode $i$ to $j$, $j$ to $k$, etc., and back to $i$ acquires the gauge-invariant phase $\Phi=\arg(g_{ij}g_{jk}\cdots g_{li})$. One can interpret $\Phi$ as the flux of an artificial magnetic field piercing the loop \cite{Koch_2010}. Thus, for $\Phi\neq 0\mod\pi$, time-reversal symmetry is broken, paving the way to a nonreciprocal response, as discussed in Refs.\cite{PhysRevApplied.4.034002, PhysRevLett.119.147703,PhysRevApplied.7.024028, PhysRevX.3.031001,Barzanjeh2017,Bernier2017,Fang2017}. In a minimal setup with three modes, the effective couplings $ g_{ij}$ can be tuned to implement a circulator \cite{Koch_2010,Ranzani2014,sliwa2015, PhysRevApplied.7.024028, Naaman_2022, kwende2023josephson} for incoming signals in resonance with the modes (see Box 1). If one of the modes is heavily damped, it can be viewed as an engineered reservoir that mediates a purely one-way interaction between the remaining modes, whenever the dissipative and coherent processes are carefully balanced using nonreciprocity~\cite{Metelmann2015, Metelmann2017}. This condition is closely related to the physics of cascaded quantum systems~\cite{PhysRevLett.70.2273, gardiner2004quantum, PhysRevLett.120.060601}, where the output field of one cavity drives another without any feedback in the reverse direction. In a larger 2D array of modes (Fig.~\ref{fig_mechanims}\textbf{e}) with lattice positions $\vec r_i$, the choice of phases $\Phi_{ij}=\arg(g_{ij})=\tfrac e\hbar\int_{\vec r_i}^{\vec r_j} \vec A_{\rm syn}\cdot d\vec r$, with $\vec A_{\rm syn}=B_{\rm syn}(-y/2,x/2,0)$, mimics the lattice dynamics of particles with charge $e$ in a perpendicular (synthetic) magnetic field of strength $B_{\rm syn}$. This configuration is particularly relevant for the simulation of magnetic effects in photonic quantum many-body systems~\cite{Ozawa2019} or the realization of topologically protected edge channels~\cite{Yao2013,Lemonde2019,Peano2015,DeBernardis2023}, e.g., for quantum communication.   

\textit{(IIb) Imposing directionality.}---Another way to employ parametric driving for breaking reciprocity is to use the external field to impose a preferred direction in a system with otherwise fully symmetric  modes~\cite{Hafezi:12, Ruesink2016, kim_non-reciprocal_2015, shen2016experimental}. This driving scheme can be implemented, for example, in devices such as toroids or ring resonators that support whispering gallery modes coupled in transmission to a waveguide driven from only one side (Fig.~\ref{fig_mechanims}\textbf{f}, top). In that case, the external driving field can mediate an effective beam-splitting interaction between the left-circulating optical mode and a (non-directional) mechanical mode of the device~\cite{Aspelmeyer2014} by amplifying the intrinsic optomechanical coupling. This amplification does not occur for the right-circulating optical mode (Fig.~\ref{fig_mechanims}\textbf{f}, bottom), rendering the scattering of any quantum signals on top of the driving field nonreciprocal. One can similarly exploit Brillouin scattering in optical waveguides to break the symmetry between signals that are co- or counter-propagating with an external driving field~\cite{kim_non-reciprocal_2015, shen2016experimental}.

\textit{(IIc) Quadrature nonreciprocity.}---Two-mode-squeezing interactions open additional possibilities for engineering devices with unconventional responses. The interference between beam-splitting and two-mode squeezing interactions generically leads to nonreciprocal, phase-sensitive amplification. In turn, phase-sensitive amplification implies a breaking of time-invariance because the amplitude of the transmitted signal is affected by a phase (or corresponding time) delay. This physical mechanism can be leveraged in an array of 1D cavities to engineer a phase-sensitive, nonreciprocal traveling wave amplifier whose amplification direction depends on the phase of the input signal~\cite{McDonald2020, wanjura_quadrature_2023, slim_optomechanical_2024, busnaina_quantum_2024}. More precisely, a signal in one quadrature is amplified with gain $G$ when propagating from the left to the right end of the device, while after exchanging the input and output ports, it is de-amplified, experiencing an effective gain of $1/G$. The amplification direction reverses if the signal is sent in the conjugate quadrature (corresponding to a phase shift of $\pi/2$). Unlike scenarios (IIa) and (IIb), this does not involve breaking time-reversal symmetry. A minimal two-mode setup to demonstrate this effect is illustrated in Fig.~\ref{fig_mechanims}\textbf{g}.

\textit{(III) Quantum states breaking time-reversal symmetry.}---In tightly confined nanophotonic waveguides, the polarisation of light is in general not transverse and linked to the direction of propagation. Such waveguides do not inherently break reciprocity, but atoms, quantum dots, or other emitters with polarization-selective transitions coupled to them could emit and absorb photons preferentially in one direction. In this case, reciprocity is not broken by the device itself, but by initializing the emitters in a specific Zeeman sublevel, e.g., by using a directional pumping field in combination with a small magnetic bias as represented in Fig.~\ref{fig_mechanims}\textbf{h}. Chiral light-matter interactions~\cite{Lodahl2017} of this type can be achieved in photonic crystal waveguides~\cite{Sollner2015}, tapered nanofibers~\cite{Petersen}, and toroidal resonators~\cite{Mitsch2014,Scheucher}, but the principle of addressing a single nonreciprocal subspace of an otherwise reciprocal system can be applied more generally. For example, by exciting only the left-circulating modes in a coupled array of toroidal resonators~\cite{Hafezi2011}, the system can mimic magnetic behavior without breaking reciprocity.  

{\it (IV) Strong few-photon nonlinearities.} The Lorentz reciprocity theorem constrains the linear response of quantum systems but not their nonlinear behavior, which becomes generically nonreciprocal in the absence of certain spatial symmetries. Thus, strong nonlinearities can induce nonreciprocal effects even in the quantum regime. A simple example is a waveguide coupled to two spatially separated two-level systems with different transition frequencies\cite{fratini2014fabry,muller2017nonreciprocal,nefedkin2022dark,hamann2018nonreciprocity}. In the ideal case of negligible loss, one two-level system fully reflects a single resonant photon, yielding zero transmission in both directions, as predicted by the linear response theory. Yet, the system exhibits strong nonreciprocity when probed with just two photons, revealing its nonlinear character.

\begin{figure*}[t!]
   \centering
\includegraphics[width=\linewidth]{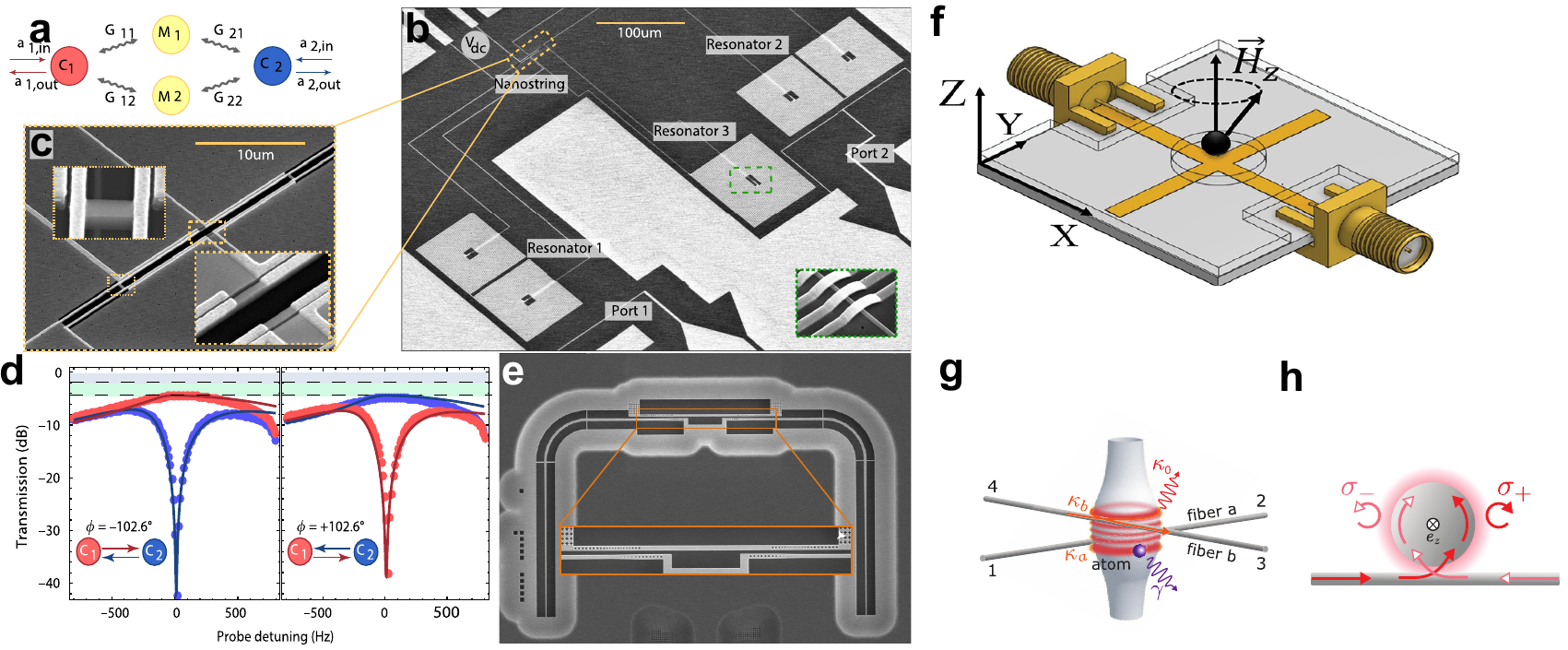}
\caption{\textbf{Experimental platforms for nonreciprocity.}
    \textbf{a}, Schematic of a chip-scale microwave isolator based on a frequency-tunable electromechanical system fabricated on a silicon-on-insulator platform. The device includes three high-impedance spiral inductors, each coupled to a dielectric nanostring mechanical resonator via vacuum-gap capacitors.
    \textbf{b}, Optical micrograph of the fabricated device showing the aluminium electrodes and nanostring resonator.
    \textbf{c}, Zoomed-in view of the mechanical nanostring, composed of two thin silicon beams suspended by symmetric tethers and patterned with four electrodes.
    \textbf{d}, Mode structure illustrating two distinct mechanical resonances that mediate interactions between the three microwave resonators. By tuning the drive configuration, directional signal transmission can be engineered across the device.
    \textbf{e}, Optomechanical implementation of nonreciprocity in a silicon optomechanical crystal circuit. Two optomechanical cavities are interconnected via optical and mechanical waveguides, forming a four-mode plaquette. Phase-coherent optical drives induce a synthetic gauge flux, enabling direction-dependent interference and yielding strong isolation with internal optical gain.
    \textbf{f}, An optomagnonic system where a YIG sphere is simultaneously coupled to both a resonator and a waveguide. The coupling to the resonator is coherent, while the waveguide provides an additional pathway for dissipative interaction with the resonator.
    \textbf{g}, A \(^{85}\text{Rb}\) atom is coupled to a whispering-gallery bottle resonator that is accessed from four ports via two tapered fibres.
    \textbf{h}, Spin--orbit locking in the evanescent field ties \(\sigma^{+}\) (\(\sigma^{-}\)) polarization to counter-clockwise (clockwise) mode.
    Panels adapted with permission from: \textbf{a}, \textbf{b}, \textbf{c}, \textbf{d}, ref.~\cite{Barzanjeh2017}, Springer Nature Ltd; \textbf{e}, ref.~\cite{Fang2017}, Springer Nature Ltd; \textbf{f}, ref.~\cite{Kim2024}, Springer Nature Ltd; \textbf{g}, \textbf{h}, ref.~\cite{Scheucher}, AAAS.}
    \label{Fig2}
\end{figure*}

\section*{Demonstrations of nonreciprocity}
Most of the mechanisms discussed above have already been demonstrated in different physical platforms. This section highlights a representative selection of those experimental approaches that illustrate how reciprocity can be broken at the quantum level and in on-chip devices. 

\textit{Optomechanical or driven systems.}---Optomechanical systems~\cite{Aspelmeyer2014,Barzanjeh2022}, where electromagnetic fields couple to mechanical motion, have proven to be versatile platforms for implementing nonreciprocity in both the microwave and optical domains. In these systems, nonreciprocal signal flow arises from parametrically driven processes, where coherent pump signals applied near the motional sideband frequencies enable interference between multiple photon--phonon pathways~\cite{Hafezi:12}. The directionality and magnitude of these interactions are governed by the relative phases of the drive tones; proper tuning of these phases allows precise control over the interference, enabling directional signal transmission~\cite{Bernier2017,Barzanjeh2017}.

This concept is illustrated in Fig.~\ref{Fig2}\textbf{a}–\textbf{c}, which shows an on-chip microwave isolator based on a frequency-tunable electromechanical system fabricated on a silicon-on-insulator platform~\cite{Barzanjeh2017}. The device consists of three high-impedance spiral inductors, each capacitively coupled to the in-plane vibrational modes of a dielectric nanostring mechanical resonator. The nanostring is composed of two thin silicon beams connected by symmetric tethers and patterned with four aluminum electrodes. These electrodes form vacuum-gap capacitors, which couple the mechanical element to three distinct microwave resonators, as shown in Fig.~\ref{Fig2}\textbf{b}–\textbf{c}. The platform supports two mechanical modes at different frequencies, enabling mediated interactions between any two or all three microwave resonators, depending on the drive configuration. As shown in Fig.~\ref{Fig2}\textbf{d}, these interactions can be engineered to yield nonreciprocal signal transmission pathways.

In the optical domain, similar nonreciprocal behavior can be achieved using fully optomechanical implementations~\cite{shen2016experimental, Ruesink2016, Fang2017}. One approach employs a silicon optomechanical crystal circuit consisting of two optomechanical cavities connected via optical and mechanical waveguides~\cite{Fang2017}, as shown in Fig.~\ref{Fig2}\textbf{e}. When coherently driven with phase-correlated optical tones, the system acquires a synthetic magnetic flux threading the four-mode plaquette. This synthetic gauge field, together with engineered mechanical dissipation, enables direction-dependent interference, with 35\,dB of isolation and 12\,dB of internal optical gain having been demonstrated in experiment. These results illustrate the effectiveness of combining synthetic magnetism and dissipation engineering to realise reconfigurable chip-scale nonreciprocal devices. By frequency-addressing multiple mechanical modes~\cite{slim_optomechanical_2024} or by employing arrays of levitated nanospheres~\cite{Rieser2022}, these techniques can be readily extended to larger lattices of coupled optical and mechanical modes.  

Beyond optomechanical systems, nonreciprocal frequency conversion between dressed optical modes has also been achieved in a compact chip-scale platform \cite{Herrmann2022}. In this work, a lithium-niobate-based device composed of three coupled racetrack resonators was modulated using dual-tone RF fields to realize mirror-symmetric frequency circulation. This approach enabled strong nonreciprocity, exceeding 40\,dB isolation, between distinct frequency modes without breaking spatial symmetry or magnetic fields, illustrating a scalable route toward fully integrated, magnet-free nonreciprocal photonic circuits.

\textit{Cavity magnons.}---Cavity magnonics has emerged as a simple approach to achieve nonreciprocity in microwave systems where large isolation and bandwidth are important and small magnetic fields can be tolerated, with immediate relevance for quantum technologies requiring compact isolators or circulators. Nonreciprocity in these systems arises from the hybridization of magnetic and photonic degrees of freedom and through two primary coupling mechanisms, coherent and dissipative. By precisely balancing these interactions, one can induce asymmetric transmission spectra resulting from interference between the coupled photon–magnon modes. This approach has enabled the realization of isolators exhibiting isolation levels greater than 50\,dB with bandwidths of several megahertz \cite{PhysRevLett.123.127202, PhysRevA.101.043842,PhysRevApplied.13.044039}. Recent experiments have confirmed the coexistence of these coherent and dissipative interactions even at millikelvin temperatures \cite{Kim2024}, making cavity magnonic devices suitable for on-chip signal routing, readout protection, and quantum-limited microwave amplification, as shown in Fig. \ref{Fig2}\textbf{f}. In Ref.~\cite{Ownes2022} it was further demonstrated that analogous schemes can be applied in 2D lattices of coupled microwave resonator arrays. Here, coherent magnon-photon interactions lift the degeneracy between left- and right-circulating cavity modes, thereby simulating the propagation of single photons in the presence of an artificial magnetic field, as illustrated in Fig.~\ref{fig_mechanims}\textbf{d}.

\textit{Saturable emitters.}---Reciprocity can be broken even in the few-photon regime by harnessing the intrinsic nonlinearity of saturable emitters coupled to a 1D waveguide. The minimal implementation is a pair of superconducting transmon qubits, which act as quantum mirrors that confine a resonant mode but with a highly nonlinear response~\cite{fratini2014fabry,muller2017nonreciprocal}. In this configuration, reciprocity is broken once spatial-inversion symmetry is lifted, and the qubits anharmonicity supplies the required nonlinearity without external pumps. Breaking this symmetry, e.g., by detuning one qubit~\cite{muller2017nonreciprocal,nefedkin2022dark}, yields a ``quantum diode'' with isolation on the order of $15$\,dB, as experimentally demonstrated in Ref.~\cite{hamann2018nonreciprocity}.  The concept extends to larger ensembles, including free-space atomic arrays~\cite{nefedkin2023nonreciprocal}, and can alternatively rely on chiral waveguides where direction-dependent coupling renders the dark state intrinsically unidirectional. Although the scattering matrix is photon-number dependent, limiting fidelity for arbitrary qubit transfer, these systems offer compact, magnet-free isolation and nonlinear optics capabilities at the quantum level.

\textit{Chiral light--matter interactions.}---As mentioned in the previous section, in photonic systems, chiral light-matter interactions can emerge in nanophotonic structures and near dielectric interfaces, where the polarization of light is linked to its direction of propagation and atoms in a magnetic field are prepared in a Zeeman sublevel. The first demonstrations of this effect were achieved with cold atoms coupled to whispering-gallery-mode (WGM) resonators and tapered fibres~\cite{ Mitsch2014, Petersen, Scheucher}. As shown in Fig.~\ref{Fig2}\textbf{g}--\textbf{h}, an atom positioned near a WGM resonator interacts asymmetrically with the clockwise and counterclockwise modes, which couple to different polarization-selective atomic transitions. This asymmetry induces direction-dependent coupling and photon loss, effectively breaking reciprocity. Moreover, the circulation direction can be reversed by preparing the atom in different Zeeman sublevels, enabling the realization of reconfigurable and programmable quantum circulators.

These systems exhibit an operational fidelity above 70\%, isolation ratios between 5 and 11\,dB, and photon survival probabilities near 73\%, with low insertion losses of approximately 1.4\,dB. Moreover, they display intrinsic nonlinearity; the routing behaviour differs for single- and two-photon inputs due to atomic saturation, allowing for the photon-number-dependent logic that is essential for photonic quantum computing. A similar mechanism for directional emission occurs in tapered nanofibers and for quantum dots embedded in chiral photonic crystal waveguides~\cite{Sollner2015}, where the tighter integration of the emitter permits directional emission with close to unit efficiency. 

\section*{Nonreciprocity for quantum technology}
Having reviewed the fundamental processes underlying nonreciprocal systems and some key realisations, we now focus on their applications in the burgeoning field of quantum technologies. We begin by exploring their role in quantum computing and simulation, focusing on quantum circuits based on mechanical systems and directional amplification. We then discuss applications in directional state transfer and quantum networks. Finally, we review recent progress in the use of nonreciprocal systems for quantum sensing, along with emerging developments in nonreciprocal energy storage in quantum devices.

\textit{Directional amplification.}---The ability to selectively amplify signals in one direction while suppressing noise and reflections in the other, i.e., known as \textit{directional amplification}, is a fundamental requirement for quantum information processing. In superconducting quantum computing, quantum-limited amplifiers \cite{PhysRevD.26.1817, RevModPhys.82.1155, Castellanos, Bergeal2010, Macklin, PhysRevApplied.22.044055, PhysRevApplied.21.064052} are crucial for high-fidelity qubit readout~\cite{Macklin, Naghiloo2019} and quantum sensing~\cite{Barzanjeh2020, GALLEGOTORROME2024100497}. However, most conventional amplifiers are reciprocal, meaning that amplified noise and backaction can propagate back to the quantum unit or node, thus degrading coherence and reducing measurement efficiency. 

\begin{figure}[t!]
\centering{\includegraphics[width=1\columnwidth]{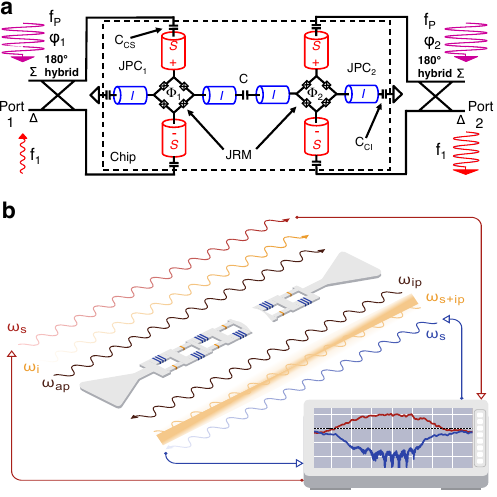}}
\caption{\textbf{|\ Nonreciprocal parametric amplification.}
\textbf{a}, Electrical schematic of the two-port device.  Input and output signals are routed through the difference (\(\Delta\)) ports of two broadband \(180^{\circ}\) hybrids that feed a pair of Josephson parametric converters (JPC\(_1\), JPC\(_2\)). Each JPC is pumped via the corresponding sum (\(\Sigma\)) port with an independent tone at the off-resonant frequency \(f_{\text{P}}\); the two pumps excite the common mode of a four-junction Josephson ring modulator (JRM). Series capacitors \(C_{\text{CS}}\) and \(C_{\text{CI}}\) couple the signal (\(S\)) and idler (\(I\)) resonators to the coplanar feed lines and to ground, setting their external quality factors. An inter-resonator capacitor \(C\) links the two \(I\) resonators and closes the interferometric loop that provides a nonreciprocal phase bias.
\textbf{b}, Operating principle of the travelling-wave parametric amplifier–isolator. A strong right-moving \textit{amplification pump} at angular frequency \(\omega_{\text{ap}}\) (brown) drives four-wave mixing, amplifying a co-propagating signal at \(\omega_{\text{i}}\) and generating an idler at \(\omega_{\text{i}}\) (yellow) to conserve energy. Concurrently, a left-moving \textit{isolation pump} at \(\omega_{\text{ip}}\) (brown) converts a counter-propagating signal at \(\omega_{\text{s}}\) (blue) up to an out-of-band frequency \(\omega_{\text{s}+\text{ip}}=\omega_{\text{s}}+\omega_{\text{ip}}\) (yellow), thereby depleting the backward wave and yielding reverse isolation. Center inset: each unit cell of the transmission line comprises three large Josephson junctions (blue) and one smaller junction (orange) embedded in an aluminium microstrip; successive cells are alternately inverted to suppress spurious harmonics. Right-hand inset: representative vector-network-analyser traces showing forward gain (red) and reverse isolation (blue) under double-pump operation.
Panels adapted with permission from: \textbf{a}, ref.~\cite{PhysRevX.3.031001}, APS; \textbf{b}, ref.~\cite{ranadive2024}.
}
\label{Fig3}
\end{figure}

Nonreciprocal amplifiers overcome this limitation by enforcing unidirectional signal transmission, thus protecting sensitive quantum states from back-propagating noise. These devices also eliminate the need for bulky and lossy external isolators, enhancing the scalability of quantum circuits. One of the earliest experimental realizations of nonreciprocal amplification in superconducting circuits utilised two coupled Josephson parametric converters (JPCs) to achieve phase-sensitive, direction-dependent gain~\cite{PhysRevX.3.031001}. This mechanism relies on parametric frequency conversion through Josephson ring modulators (JRMs), enabling back-to-back JPC operation with tailored nonreciprocity, as shown in Fig.~\ref{Fig3}\textbf{a}. This system achieves forward amplification of 15\,dB gain across a 10–30\,MHz bandwidth, suppresses reverse transmission by 6\,dB, breaks reciprocity without ferrite-based components, exhibits near-quantum-limited performance with added noise below 1.5 photons~\cite{PhysRevLett.112.167701}.

Broadly speaking, directional amplifiers fall into two categories, with resonance-based narrowband amplifiers~\cite{PhysRevX.3.031001, PhysRevLett.112.167701, sliwa2015, PhysRevApplied.7.024028} on the one hand, and broadband travelling-wave parametric amplifiers (TWPAs)~\cite{9134828, Macklin, Esposito} on the other. Resonance-based directional amplifiers can provide strong unidirectional gain. Still, their operational bandwidths, for amplification as well as isolation, are fundamentally limited by the narrow frequency response of the resonant elements. By contrast, TWPAs can deliver high gain over much broader bandwidths, making them highly attractive for applications in quantum signal processing. However, standard TWPAs lack built-in isolation, allowing backward-propagating signals to reach sensitive components and degrade system performance. A recent development introduced a superconducting TWPA architecture that incorporates isolation directly into its design~\cite{ranadive2024}. This Josephson-based TWPA isolator eliminates the need for external isolators while maintaining near-quantum-limited noise performance, marking an advancement toward scalable quantum amplification systems. Figure~\ref{Fig3}\textbf{b} shows the schematic of the TWPA isolator. The design utilizes third-order nonlinear wave mixing to amplify forward-propagating signals, while second-order nonlinear processes are employed to frequency-upconvert and reject backward-propagating signals. This hybrid mechanism enables robust nonreciprocity, where forward signals are amplified with 20\,dB gain, while reverse signals are suppressed by 30\,dB. Importantly, the device achieves this performance over a 500\,MHz bandwidth, significantly surpassing the limitations of earlier resonance-based superconducting amplifiers.

\textit{Topological amplification.}---The application of topological concepts has been fruitful in discovering new types of nonreciprocal amplifiers and in providing new physical insights into nonreciprocal amplification. Ref.\cite{Peano_2016} showed how to take advantage of the lack of backscattering in the topological edge states of 2D Chern photonic insulator\cite{Ozawa2019} to create a quantum-limited amplifier with built-in reverse isolation that is particularly robust against disorder. This concept was experimentally demonstrated in a 2D array of ring resonators by Mittal, \textit{et al.}\cite{Mittal2018}.  Furthermore, several theoretical studies \cite{Porras2019, Wanjura2020, Wanjura2021} have demonstrated an intriguing link between non-Hermitian topology and directional amplification in traveling-wave amplifiers based on one-dimensional cavity arrays. 
More specifically, directional amplification is in one-to-one correspondence with the nontrivial topology of the bulk non-Hermitian Hamiltonian (or dynamical matrix).

\textit{Directional state transfer.}---Quantum networks enable the transmission of quantum signals between physically distant nodes, which is usually achieved through the controlled emission and successive reabsorption of individual optical or microwave photons~\cite{Cirac1997}. Apart from the ability to tune the qubit–waveguide coupling in time, this requires directional photon emission to ensure that the complete photonic quantum state reaches the targeted receiver qubit. These quantum routing capabilities---essential for the scalability of any quantum information processing architecture---are thus fundamentally linked to the implementation of nonreciprocal network topologies.

\begin{figure}[t!]
\centering{\includegraphics[width=1\columnwidth]{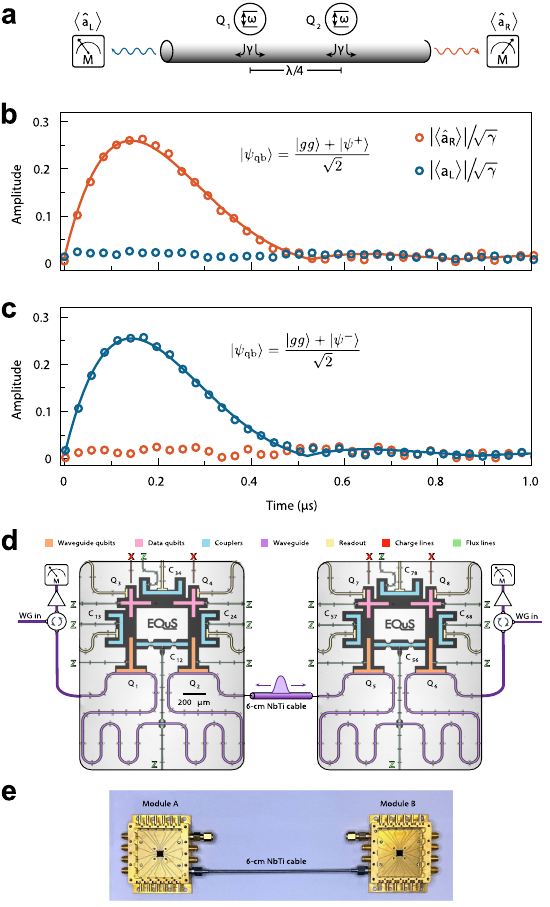}}
\caption{\textbf{Platforms for directional state transfer.}
\textbf{a}, Two transmon qubits coupled to a waveguide and separated by $\lambda/4$ form a ``chiral'' emitter; a pulse sequence entangles ancillary data qubits and maps the excitation onto a travelling photon.
\textbf{b}, \textbf{c}, Time-resolved field amplitudes measured at the two ends of the waveguide demonstrate right- or left-propagating photon emission, as selected by the initial two-qubit phase.
\textbf{d}, Optical micrographs of two identical superconducting modules, each capable of emitting and absorbing directional photons and linked by a NbTi cable.
\textbf{e}, Photograph of the packaged modules constituting the cryogenic interconnect.
Panels adapted with permission from: \textbf{a, b, c}, ref.~\cite{Kannan2023}, Springer Nature Ltd; \textbf{d, e}, ref.~\cite{Almanakly2025}, Springer Nature Ltd.}
\label{Fig4}
\end{figure}

A paradigmatic platform where directional state transfer techniques are currently being developed is superconducting qubits coupled via microwave transmission lines. In these systems, directional qubit-photon interfaces can be realized by exploiting interference effects between distributed coupling points. In the basic configuration depicted in Fig.~\ref{Fig4}\textbf{a}, two qubits that are separated by a distance of $d=\lambda/4$, where $\lambda$ is the resonant photon wavelength, decay into a common microwave channel. Preparation of the qubits in the states $|\psi^\pm\rangle=\bigl(|eg\rangle\pm e^{i\pi/4}|ge\rangle\bigr)\big/\sqrt{2}$, with $\bigl|g\rangle$ and $\bigl|e\rangle$ being the two states of each qubit, gives rise to an additional propagation phase $\pm\tfrac\pi2$, which leads to constructive or destructive interference depending on the direction of emission~\cite{Gheeraert2020,Guimond2020}. Based on this principle, emission fidelities exceeding 95\% have been demonstrated in a first proof-of-principle experiment~\cite{Kannan2023}, as shown in Fig.~\ref{Fig4}\textbf{b}--\textbf{c}. A similar scheme has been implemented by Joshi, \textit{et al.}~\cite{Joshi2023} with a single qubit that is connected to the waveguide at two separate coupling points. In this case, the directionality is imposed by adjusting the phase in one of the legs through a tunable Josephson coupler. Variants of this idea have also been discussed in the context of giant atoms~\cite{Wang2022,Soro2022, Yao_giant_2022}, where different coupling patterns offer additional functionalities and flexibility.  

Going beyond a single node, deterministic entanglement between two remote superconducting modules was achieved by exploiting chiral photon exchange through a shared waveguide~\cite{Almanakly2025}. As depicted in Fig.~\ref{Fig4}\textbf{d}--\textbf{e}, each module includes data qubits (used for entanglement preparation and storage) and waveguide qubits (used for photon emission and absorption), and parametric modulation of the couplers enables wavepacket shaping. This approach yielded remote entanglement in the form of a four-qubit W-state, achieving fidelity exceeding 62\% in both directions. Importantly, the method is independent of the physical separation between modules, offering intrinsic scalability for quantum networks with flexible connectivity and robust entanglement distribution.

\textit{Nonreciprocal quantum sensors.}---Quantum sensing aims to surpass the limits of classical sensing by using quantum resources to achieve higher precision in detecting weak signals. A central challenge in such measurements is to maximise sensitivity while minimising the impact of noise. Nonreciprocity provides a powerful way to address this challenge~\cite{Lau2018}. In conventional reciprocal systems, signals and backaction noise share the same propagation paths, making it difficult to amplify or detect a signal without amplifying the accompanying noise. Nonreciprocal systems make it possible to break this symmetry, allowing signals to be directed along one path while backaction is suppressed or routed elsewhere. As a result, nonreciprocal architectures can significantly boost the performance of quantum sensors, achieving faster measurement rates, higher signal-to-noise ratios, and, ultimately, more precise detection of weak perturbations than is possible with reciprocal systems.

\begin{figure}[t!]
\centering{\includegraphics[width=0.9\columnwidth]{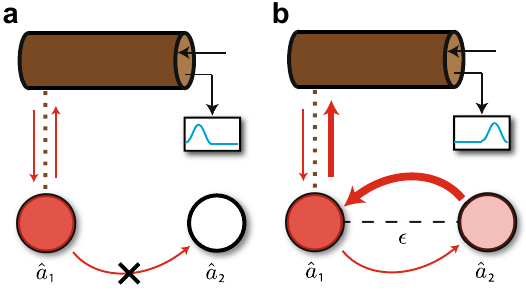}}
\caption{\textbf{Signal enhancement via nonreciprocity.} Arrows indicate the direction and strength of coupling between two modes.
\textbf{a}, In the unperturbed case, strong unidirectional coupling from mode 2 to mode 1 remains inaccessible due to perfect nonreciprocity, and mode 2 is not excited by a drive on mode 1.
\textbf{b}, A small perturbation enables photons to tunnel from mode 1 to mode 2, which then return via the nonreciprocal path with amplitude $J$, leading to effective amplification and enhanced signal readout.
Panels adapted with permission from: \textbf{a, b}, ref.~\cite{McDonald2020}, Springer Nature Ltd.}
\label{FigSensing}
\end{figure}

Much recent work on enhanced sensing focuses on exceptional points \cite{PhysRevA.93.033809, Hodaei2017, Chen2017, Mao}, which are spectral degeneracies where the eigenvalues and eigenvectors of a system coalesce. This degeneracy can be achieved in a two-mode system, e.g., by tuning coupling and dissipation rates. Near these exceptional points, small parameter changes can induce disproportionately large shifts in resonance frequencies, suggesting enhanced sensitivity. However, when (quantum) noise and realistic measurement protocols are taken into account, it becomes evident that exceptional points do not enhance the fundamental measurement rate \cite{Langbein2018,Duggan2022,ding2023}; any gain in signal strength is accompanied by an increase in noise, and the overall signal-to-noise ratio remains bounded by limits also achievable in simpler systems. A distinct advantage in signal-to-noise ratio does arise when the two modes are coupled asymmetrically beyond what is possible in a reciprocal system (see Fig~\ref{FigSensing}). This enhancement does not come from spectral singularities, but from an increase in the measurement rate even when the total intracavity photon number, a key constraint in quantum systems, is fixed.

In more complex architectures, such as parametric bosonic lattices, the benefits of nonreciprocity become even more pronounced \cite{McDonald2020}. By engineering asymmetric coupling, e.g., through parametric drives, these systems support chiral propagation of different quadrature components. When a localised perturbation couples two oppositely chiral modes, it creates a round-trip signal path; a probe signal travels through one directionally amplified channel, interacts with the parameter to be estimated, and is scattered back along another amplified path. This configuration results in a homodyne signal whose amplitude grows exponentially with system size, while the quantum noise remains bounded. In this regime, the quantum Fisher information per photon, a fundamental metric of metrological performance, also scales exponentially, representing a significant breakthrough in quantum-limited sensing.

\section*{Observations, outlook, and outstanding challenges}
Nonreciprocity and asymmetric transmission have become indispensable tools in the quantum engineering toolbox, enabling directional control of signal flow, isolation from noise, and enhanced measurement capabilities. Recent advances in superconducting circuits, nanophotonics, and hybrid optomechanical systems have demonstrated that nonreciprocal behaviour can now be realised in compact and often magnet-free devices that are fully compatible with on-chip quantum architectures. This represents a fundamental shift away from traditional bulky magneto-optic components and paves the way for scalable quantum sensing, communication, and information processing technologies. 

As quantum platforms continue to evolve, nonreciprocal elements are expected to play an increasingly central role, not merely as auxiliary devices for protection or filtering, but as core components that determine how quantum information is transmitted, measured, and stabilized. Nonreciprocity offers powerful routes to implement deterministic state transfer, unidirectional photon--matter interactions, and cascaded quantum systems \cite{gardiner2004quantum}. These capabilities are essential for modular quantum computing \cite{Wang24}, distributed entanglement generation, and real-time quantum feedback and control \cite{PhysRevA.95.013837}. In addition to its roles in communication and measurement, nonreciprocity has also shown promise in quantum energy transport and storage. Recent works \cite{Ahmadi2024} have demonstrated that nonreciprocity in quantum batteries can significantly enhance energy accumulation by directing energy flow and suppressing unwanted dissipation. These insights suggest broader implications for quantum sensors and processors that rely on energy-efficient readout mechanisms and power management architectures.

Looking ahead, several directions are especially promising. Hybrid approaches that combine synthetic gauge fields, optomechanical coupling, and time-modulated interactions offer opportunities for programmable, reconfigurable nonreciprocity with a large dynamic range. Applications include entangled photon routers, topological photonic lattices, non-Hermitian devices for quantum simulation and precision metrology, and energy-efficient quantum hardware enabled by directional control of energy flow. Nonetheless, important challenges remain that must be addressed to realise the potential of nonreciprocal quantum technologies. Achieving strong nonreciprocity at the few- or single-photon level without introducing excess noise or loss remains a central obstacle. Current approaches are limited by narrow bandwidths, susceptibility to phase instabilities, and complex multi-tone parametric drives. Ensuring long-term stability, thermal robustness, and seamless integration with cryogenic quantum hardware is another pressing concern.

Future progress will require coordinated advances in theory, materials, and device engineering. In particular, the development of low-loss, broadband, and dynamically tunable nonreciprocal components, compatible with superconducting, photonic, and hybrid quantum systems, will be crucial. If these technical barriers can be overcome, nonreciprocity may unlock entirely new paradigms in quantum error correction, fault-tolerant architectures, energy-aware computing, and distributed quantum networks, making it a cornerstone of next-generation quantum technologies.
\acknowledgements{%
S.B.\ acknowledges funding by the Natural Sciences and Engineering Research Council of Canada (NSERC) through its Discovery Grant, funding and advisory support provided by Alberta Innovates through the Accelerating Innovations into CarE (AICE) – Concepts Program, and support from Alberta Innovates and NSERC through Advance Grant. A.X.\ acknowledges support from the European Union’s Horizon Europe research and innovation programme through projects Quantum Secure Networks Partnership (QSNP; grant agreement No 101114043) and QUantum DevIces and subsystems for Communications in spacE (QUDICE; grant agreement No 101082596), as well as the Ministry for Education, Sports, Youths, Research and Innovation of the Government of Malta through its participation in the QuantERA ERA-NET Cofund in Quantum Technologies (project MQSens) implemented within the European Union’s Horizon 2020 research and innovation programme. This research is part of the Munich Quantum Valley, which is supported by the Bavarian state government with funds from the Hightech Agenda Bayern Plus.}

\end{document}